\documentclass[12pt]{article}

\usepackage{bm}

\begin{document}

\title{{\bf Incompatibility of the Dirac-like Field Operators with the Majorana Anzatzen}}

\author{{\bf Valeriy V. Dvoeglazov}\\ UAF, Universidad Aut\'onoma de Zacatecas, M\'exico\\E-mail: valeri@fisica.uaz.edu.mx}

\date{\empty}

\maketitle

\begin{abstract}
In the present article we investigate the spin-1/2 and spin-1 cases in different bases. Next, we look for relations 
with the Majorana-like field operator. We show explicitly incompatibility of the Majorana anzatzen with the Dirac-like field operators in both the original Majorana theory and its generalizations. Several explicit examples are presented for higher spins too. It seems that the calculations in the helicity basis only give mathematically and physically reasonable results.
\end{abstract}

\section{Introduction.}

In Refs.~\cite{DV1}-\cite{DVCONF2} we considered the procedure of construction of the field operators {\it ab initio} 
(including for neutral particles). The Bogoliubov-Shirkov method has been used.

In the present article we investigate the spin-1/2 and spin-1 cases in different bases. The Majorana theory of the neutral particles is well known~\cite{Majorana}. We look for relations of the Dirac-like field operator 
to the Majorana-like field operator. It seems that the calculations in the helicity basis give mathematically and physically reasonable results.


\section{The Spin-1/2.}

The Dirac equation is:
\begin{equation}
[i\gamma^\mu \partial_\mu -mc/\hbar]\Psi (x) =0\,,\label{Dirac}
\end{equation}
$\mu =0,1,2,3$.
The most known methods of its derivation are~\cite{Dirac,Sakurai,Ryder}:
\begin{itemize}
  \item the Dirac one (the Hamiltonian should be linear in $\partial/\partial x^i$, and be compatible with $E_p^2 -{\bf p}^2 c^2 =m^2 c^4$);
  \item the Sakurai one (based on the equation $(E_p- c{\bf \sigma} \cdot {\bf p}) (E_p+ c{\bf \sigma} \cdot {\bf p}) \phi 
=m^2 c^4 \phi$, $\phi$ is the 2-component spinor);
  \item the Ryder one (the relation between  2-spinors at rest is $\phi_R ({\bf 0}) = \pm \phi_L ({\bf 0})$, and boosts).
\end{itemize}
The $\gamma^\mu$ are the Clifford algebra matrices 
\begin{equation}
\gamma^\mu \gamma^\nu +\gamma^\nu \gamma^\mu = 2g^{\mu\nu}\,,
\end{equation}
$g^{\mu\nu}$ is the metric tensor.
Usually, everybody uses the following definition of the field operator~\cite{Itzyk} in the pseudo-Euclidean metrics:
\begin{equation}
\Psi (x) = \frac{1}{(2\pi)^3}\sum_h \int \frac{d^3 {\bf p}}{2E_p} [ u_h ({\bf p}) a_h ({\bf p}) e^{-ip_\mu\cdot x^\mu}
+ v_h ({\bf p}) b_h^\dagger ({\bf p}) e^{+ip_\mu\cdot x^\mu}]\,,
\end{equation}
as given {\it ab initio}.
After actions of the Dirac operator on  $\exp (\mp ip_\mu x^\mu)$ the 4-spinors ( $u-$ and $v-$ ) 
satisfy the momentum-space equations: $(\hat p - m) u_h (p) =0$ and $(\hat p + m) v_h (p) =0$, respectively; the $h$ is 
the polarization index. It is easy to prove from the characteristic equations
$Det (\hat p \mp m) =(p_0^2 -{\bf p}^2 -m^2)^2= 0$ that the solutions should satisfy the energy-momentum relations $p_0= \pm E_p =\pm \sqrt{{\bf p}^2 +m^2}$ for both $u-$ and $v-$ solutions.

The general scheme of construction of the field operator has been presented in~\cite{Bogoliubov}. In the case of
the $(1/2,0)\oplus (0,1/2)$ representation we have:
\begin{equation}
\Psi (x) = \frac{1}{(2\pi)^{3}}\int dp  e^{ip\cdot x}\tilde \Psi (p)\,.
\end{equation}
We know the condition of the mass shell: $(p^2 -m^2) \tilde \Psi (p) =0$.
Thus, $\tilde \Psi (p) = \delta (p^2 - m^2) \Psi (p)$.
Next, 
\begin{eqnarray}
\Psi (x) &=& \frac{1}{(2\pi)^{3}} \int d p \, e^{ip\cdot x}\delta (p^2 -m^2) (\theta (p_0)+ 
\theta (-p_0)) 
\Psi (p) =\\
&=& \frac{1}{(2\pi)^{3}} \int d p \left [ e^{ip\cdot x} \delta (p^2 -m^2) \Psi^+ (p)
+ e^{-ip\cdot x} \delta (p^2 -m^2) \Psi^- (p)\right ],\nonumber
\end{eqnarray}
where
\begin{equation}
\Psi^+ (p) =\theta (p_0) \Psi (p)\,,\mbox{and}\,\, \Psi^- (p)=\theta (p_0) \Psi (-p)\,.
\end{equation}
\begin{eqnarray}
\Psi^+ (x) &=& \frac{1}{(2\pi)^{3}}\int \frac{d^3 {\bf p}}{2E_p}  e^{+ip\cdot x} \Psi^+ (p)\,,\\
\Psi^- (x) &=& \frac{1}{(2\pi)^{3}}\int \frac{d^3 {\bf p}}{2E_p}  e^{-ip\cdot x} \Psi^- (p)\,.
\end{eqnarray}
We adjust the notation to the modern one:
\begin{eqnarray}
&&\Psi (x) = {1\over (2\pi)^3} \int d^4 p \,\delta (p^2 -m^2) e^{-ip\cdot x}
\Psi (p) =\nonumber\\
&=& {1\over (2\pi)^3} \sum_{h}^{}\int d^4 p \, \delta (p_0^2 -E_p^2) e^{-ip\cdot x}
u_h (p_0, {\bf p}) a_h (p_0, {\bf p}) =\nonumber\\
&=&{1\over (2\pi)^3} \int {d^4 p \over 2E_p} [\delta (p_0 -E_p) +\delta (p_0 +E_p) ] 
[\theta (p_0) +\theta (-p_0) ] e^{-ip\cdot x} \times \nonumber\\
&\times&\sum_{h}^{} u_h (p) a_h (p)= {1\over (2\pi)^3} \sum_h^{} \int {d^4 p \over 2E_p}
[\delta (p_0 -E_p) +\delta (p_0 +E_p) ] \times\label{fo}\\
&\times&\left [\theta (p_0) u_h (p) a_h (p) e^{-ip\cdot x}  +
\theta (p_0) u_h (-p) a_h (-p) e^{+ip\cdot x} \right ] =\nonumber\\
&=& {1\over (2\pi)^3} \sum_h^{} \int {d^3 {\bf p} \over 2E_p} \theta(p_0)  
\left [ u_h (p) a_h (p)\vert_{p_0=E_p} e^{-i(E_p t-{\bf p}\cdot {\bf x})}+ \right. \nonumber\\
&+& \left. u_h (-p) a_h (-p)\vert_{p_0=E_p} e^{+i (E_p t- {\bf p}\cdot {\bf x})} 
\right ]\nonumber
\end{eqnarray}
During the calculations we had to represent $1=\theta (p_0) +\theta (-p_0)$ above
in order to get positive- and negative-frequency parts.\footnote{See Ref.~[5b]
for discussion.} Moreover, we did not yet assumed, which equation this
field operator  (namely, the $u-$ spinor) satisfies, with negative- or positive- mass and/or $p^0= \pm E_p$.

In general we should transform $u_h (-p)$ to the $v_h (p)$. The procedure is the following one~\cite{DV1,DV2}.

\subsection{The Standard Basis.}

The explicit forms of the 4-spinors are:
\begin{eqnarray}
u_\sigma ({\bf p})= \frac{N_\sigma^+}{2\sqrt{m (E_p +m)}}\pmatrix{[E_p+m+{\bm \sigma}\cdot {\bf p}]\phi_\sigma ({\bf 0})\cr
[E_p+m-{\bm \sigma}\cdot {\bf p}]\chi_\sigma ({\bf 0})\cr}\,,\quad v_\sigma ({\bf p}) =\gamma^5  u_\sigma ({\bf p})\,,\label{s1a}
\end{eqnarray}
in the spinorial basis ($\phi_\uparrow ({\bf 0}) =\chi_\uparrow  ({\bf 0})= \pmatrix{1\cr 0\cr}$ 
and $\phi_\downarrow ({\bf 0}) =\chi_\downarrow  ({\bf 0})= \pmatrix{0\cr 1\cr}$). The transformation to  
the standard basis is produced with 
the $(\gamma^5 +\gamma^0)/\sqrt{2}$ matrix. The normalizations, projection operators, 
propagators, dynamical invariants {\it etc} have been given in~\cite{Ryder}, for example.

In the Dirac case we should assume the following relation in the field operator (\ref{fo}):
\begin{equation}
\sum_{h=\pm 1/2}^{} v_h (p) b_h^\dagger (p) = \sum_{h=\pm 1/2}^{} u_h (-p) a_h (-p)\,,\label{dcop}
\end{equation}
which is compatible with the ``hole" theory and the Feynman-Stueckelberg interpretation.
We know that~\cite{Ryder}
\begin{eqnarray}
\bar u_\mu (p) u_\lambda (p) &=& +m \delta_{\mu\lambda}\,,\\
\bar u_\mu (p) u_\lambda (-p) &=& 0\,,\\
\bar v_\mu (p) v_\lambda (p) &=& -m \delta_{\mu\lambda}\,,\\
\bar v_\mu (p) u_\lambda (p) &=& 0\,,
\end{eqnarray}
$\mu$, $\lambda$ are now the polarization indices.
However, we need $\Lambda_{\mu\lambda} (p) = \bar v_\mu (p) u_\lambda (-p)$.
By direct calculations,  we find
\begin{equation}
-mb_\mu^\dagger (p) = \sum_{\lambda}^{} \Lambda_{\mu\lambda} (p) a_\lambda (-p)\,.
\end{equation}
Hence, $\Lambda_{\mu\lambda} = - im ({\bm \sigma}\cdot {\bf n})_{\mu\lambda}$, ${\bf n} = {\bf p}/\vert{\bf p}\vert$, 
and 
\begin{equation}
b_\mu^\dagger (p) = + i\sum_\lambda ({\bm\sigma}\cdot {\bf n})_{\mu\lambda} a_\lambda (-p)\,.\label{co}
\end{equation}
Multiplying (\ref{dcop}) by $\bar u_\mu (-p)$ we obtain
\begin{equation}
a_\mu (-p) = -i \sum_{\lambda} ({\bm \sigma} \cdot {\bf n})_{\mu\lambda} b_\lambda^\dagger (p)\,.\label{ao}
\end{equation}
The equations are self-consistent.

\subsection{The Helicity Basis.}

The 2-eigenspinors of the helicity operator 
\begin{eqnarray}
{1\over 2} {\bm \sigma}\cdot\widehat
{\bf p} = {1\over 2} \pmatrix{\cos\theta & \sin\theta e^{-i\phi}\cr
\sin\theta e^{+i\phi} & - \cos\theta\cr}
\end{eqnarray}
can be defined as follows~\cite{Var,Dv1,DVIJTP}:
\begin{eqnarray}
\phi_{{1\over 2}\uparrow}=\pmatrix{\cos{\theta \over 2} e^{-i\phi/2}\cr
\sin{\theta \over 2} e^{+i\phi/2}\cr}\,,\quad
\phi_{{1\over 2}\downarrow}=\pmatrix{\sin{\theta \over 2} e^{-i\phi/2}\cr
-\cos{\theta \over 2} e^{+i\phi/2}\cr}\,,\quad\label{ds}
\end{eqnarray}
for $\pm 1/2$ eigenvalues, respectively.

We start from the Klein-Gordon equation, generalized for
describing the spin-1/2  particles (i.~e., two degrees
of freedom); $c=\hbar=1$:
\begin{equation}
(p_0+{\bm \sigma}\cdot {\bf p}) (p_0- {\bm \sigma}\cdot {\bf p}) \phi
= m^2 \phi\,.\label{de}
\end{equation}
It can be re-written in the form of the set of two first-order equations
for 2-spinors. We observe at the same time that they may be chosen
as eigenstates of the helicity operator which present
in (\ref{de}):\footnote{This opposes to the choice of the basis of the subsection
(2.1), where 4-spinors are the eigenstates of the parity operator, cf.~\cite{BLP}.}
\begin{eqnarray}
(p_0-({\bm\sigma}\cdot {\bf p})) \phi_\uparrow &=& (p_0-p) \phi_\uparrow
=m\chi_\uparrow \,,\\
(p_0+({\bm\sigma}\cdot {\bf p})) \chi_\uparrow &=& (p_0+p) \chi_\uparrow
=m\phi_\uparrow \,,\\
(p_0-({\bm\sigma}\cdot {\bf p})) \phi_\downarrow &=& (p_0+p) \phi_\downarrow
=m\chi_\downarrow\,, \\
(p_0+({\bm\sigma}\cdot {\bf p})) \chi_\downarrow &=& (p_0-p) \chi_\downarrow
=m\phi_\downarrow \,.
\end{eqnarray}
If the $\phi_{\uparrow\downarrow}$ spinors are defined by the equation (\ref{ds}) then we
can construct the corresponding $u-$ and $v-$ 4-spinors:\footnote{Alternatively, $\uparrow\downarrow$ 
may refer to the chiral helicity
eigenstates, e.g. $u_\eta ={1\over \sqrt{2}} \pmatrix{N \phi_\eta\cr
N^{-1} \phi_{-\eta}\cr}$, see next sections and cf.~\cite{extra,DVA1995}.}
\begin{eqnarray}
u_\uparrow ({\bf p}) &=&
N_\uparrow^+ \pmatrix{\phi_\uparrow\cr {E_p-p\over m}\phi_\uparrow\cr} =
{1\over \sqrt{2}}\pmatrix{\sqrt{{E_p+p\over m}} \phi_\uparrow\cr
\sqrt{{m\over E_p+p}} \phi_\uparrow\cr}\,,\nonumber\\
u_\downarrow ({\bf p}) &=& N_\downarrow^+ \pmatrix{\phi_\downarrow\cr
{E_p+p\over m}\phi_\downarrow\cr} = {1\over
\sqrt{2}}\pmatrix{\sqrt{{m\over E_p+p}} \phi_\downarrow\cr \sqrt{{E_p+p\over
m}} \phi_\downarrow\cr}\,,\label{s1}\\
v_\uparrow ({\bf p}) &=& N_\uparrow^- \pmatrix{\phi_\uparrow\cr
-{E_p-p\over m}\phi_\uparrow\cr} = {1\over \sqrt{2}}\pmatrix{\sqrt{{E_p+p\over
m}} \phi_\uparrow\cr
-\sqrt{{m\over E_p+p}} \phi_\uparrow\cr}\,, \nonumber\\
v_\downarrow ({\bf p}) &=& N_\downarrow^- \pmatrix{\phi_\downarrow\cr
-{E_p+p\over m}\phi_\downarrow\cr} = {1\over
\sqrt{2}}\pmatrix{\sqrt{{m\over E_p+p}} \phi_\downarrow\cr -\sqrt{{E_p+p\over
m}} \phi_\downarrow\cr}\,,\label{s2}
\end{eqnarray}
where the normalization to the unit
($\pm 1$) was used:\footnote{Of course, there are no any mathematical
difficulties
to change it  to the normalization to $\pm m$, which may be more convenient
for the study of the massless limit.}
\begin{eqnarray}
\bar u_h ({\bf p}) u_{h^\prime} ({\bf p}) &=& \delta_{hh^\prime}\,,
\bar v_h ({\bf p}) v_{h^\prime} ({\bf p}) = -\delta_{hh^\prime}\,,\\
\bar u_h ({\bf p}) v_{h^\prime} ({\bf p}) &=& 0 =
\bar v_h ({\bf p}) u_{h^\prime} ({\bf p})
\end{eqnarray}

We define the field operator as follows:
\begin{equation}
\Psi (x^\mu) = \sum_h \int \frac{d^3 {\bf p}}{(2\pi)^3}
\frac{\sqrt{m}}{2E_p} [ u_h ({\bf p}) a_h ({\bf p}) e^{-ip_\mu x^\mu} +v_h
({\bf p}) b^\dagger_h ({\bf p}) e^{+ip_\mu x^\mu} ]\,.
\end{equation}
The commutation
relations are assumed to be the standard
ones~\cite{Bogoliubov,Itzyk,Wein,Greinb}\footnote{The only possible changes
may be related to different forms of normalization of 4-spinors,
which would have influence on the factor before $\delta$-function.}
(compare  with~\cite{Tokuoka})
\begin{eqnarray}
&&\left [a_h ({\bf p}),
a_{h^\prime}^\dagger ({\bf k})
\right ]_+ = 2E_p \delta^{(3)} ({\bf p}-{\bf k})
\delta_{hh^\prime}\,,
\left [a_h ({\bf p}),
a_{h^\prime} ({\bf k})\right ]_+ = 0 =
\left [a_h^\dagger ({\bf p}),
a_{h^\prime}^\dagger ({\bf k})\right ]_+\nonumber\\
&&\\
&&\left [a_h ({\bf p}),
b_{h^\prime}^\dagger ({\bf k})\right ]_+ = 0 =
\left [b_h ({\bf p}),
a_{h^\prime}^\dagger ({\bf k})\right ]_+,\\
&&\left [b_h ({\bf p}),
b_{h^\prime}^\dagger ({\bf k})
\right ]_+ = 2E_p \delta^{(3)} ({\bf p}-{\bf k})
\delta_{hh^\prime}\,,
\left [b_h ({\bf p}),
b_{h^\prime} ({\bf k})\right ]_+ = 0 =
\left [b_h^\dagger ({\bf p}),
b_{h^\prime}^\dagger ({\bf k})\right ]_+
\end{eqnarray}
Other details of the helicity basis are given in Refs.~\cite{GR,DVIJTP}.

However, in this helicity case we have:
\begin{equation}
\Lambda_{hh^\prime} (p) = \bar v_h (p) u_{h^\prime} (-p)=
i\sigma^y_{hh^\prime}\,.
\end{equation}
So, someone  may argue that we should introduce the creation operators by hand in every basis.

\subsection{Application of the Majorana anzatzen.}

It is well known that  ``{\it particle=antiparticle}" in the Majorana theory~\cite{Majorana}. So, in the language of the quantum field theory we should have 
\begin{equation}
b_\mu (E_p, {\bf p}) = e^{i\varphi} a_\mu (E_p, {\bf p})\,.\label{ma}
\end{equation} 
Usually, different authors use $\varphi = 0, \pm \pi/2$ depending on the metrics and on the forms of the 4-spinors and commutation relations.

So, on using (\ref{co}) and the above-mentioned postulate we come to:
\begin{equation}
a_\mu^\dagger (p) = +i e^{i\varphi} ({\bm \sigma}\cdot {\bf n})_{\mu\lambda} a_\lambda (-p)\,.\label{ma1}
\end{equation}
On the other hand, on using (\ref{ao}) we make the substitutions $E_p \rightarrow -E_p$, ${\bf p} 
\rightarrow -{\bf p}$ to obtain 
\begin{equation}
a_\mu (p) = +i ({\bm \sigma}\cdot {\bf n})_{\mu\lambda} b_\lambda^\dagger (-p)\,.\label{ma2}
\end{equation}
The totally reflected (\ref{ma}) is $b_\mu (-E_p, -{\bf p}) = e^{i\varphi} a_\mu (-E_p, -{\bf p})$.\footnote{We should have the same
contradiction even if $\varphi \rightarrow \alpha$.} Thus,
\begin{equation}
b_\mu^\dagger (- p) = e^{-i\varphi} a_\mu^\dagger (- p)\,.
\end{equation}
Combining with (\ref{ma2}),  we come to
\begin{equation}
a_\mu (p) = +i e^{-i\varphi}({\bm \sigma}\cdot {\bf n})_{\mu\lambda} a_\lambda^\dagger (-p)\,,
\end{equation}
and 
\begin{equation}
a_\mu^\dagger (p) = -i e^{i\varphi}({\bm \sigma}^\ast \cdot {\bf n})_{\mu\lambda} a_\lambda (-p)\,.
\end{equation}
This contradicts with the equation (\ref{ma1}) unless we have the preferred axis in every inertial system.

Next, we can use another Majorana anzatz $\Psi = \pm e^{i\alpha} \Psi^{c}$ with usual definitions
\begin{eqnarray}
{\cal C} =e^{i\vartheta_c}\pmatrix{0&i\Theta\cr -i\Theta & 0\cr} {\cal K}\,,\quad
\Theta = \pmatrix{0&-1\cr 1&0\cr}\,.\label{cco}
\end{eqnarray}
Thus, on using $Cu_\uparrow^\ast ({\bf p}) =iv_\downarrow ({\bf p})$, $Cu_\downarrow^\ast ({\bf p}) =-iv_\uparrow ({\bf p})$ we come to other relations
between creation/annihilation operators 
\begin{eqnarray}
a_\uparrow^\dagger  ({\bf p}) &=& \mp i e^{-i\alpha} b_\downarrow^\dagger ({\bf p})\,,\\
a_\downarrow^\dagger  ({\bf p}) &=& \pm i e^{-i\alpha} b_\uparrow^\dagger ({\bf p})\,,
\end{eqnarray}
which may be used instead of (\ref{ma}).
Due to the possible signs $\pm$ the number of the corresponding states is the same as in the Dirac case that permits us to have the complete system 
of the Fock states over the $(1/2,0)\oplus (0,1/2)$ representation space in the mathematical sense.\footnote{Please note that the phase factors may have physical significance in quantum field theories as opposed to the textbook nonrelativistic quantum mechanics, as was discussed recently by several authors.}
However, in this case we deal with the self/anti-self charge conjugate quantum field operator instead of the self/anti-self charge conjugate quantum states. Please remember that it is the latter that answer for the neutral particles; the quantum field operator 
contains operators for more than one state, which may be either electrically neutral or charged.

We conclude that something is missed in the foundations of both the original Majorana theory and its generalizations
in the $(1/2,0)\oplus (0,1/2)$ representation.

\subsection{Self/Anti-self Charge Conjugate States.}

We re-write the charge conjugation operator (\ref{cco}) to the form:
\begin{equation}
{\cal C}= e^{i\vartheta_c} \pmatrix{0&0&0&-i\cr
0&0&i&0\cr
0&i&0&0\cr
-i&0&0&0\cr} {\cal K} = -e^{i\vartheta_c} \gamma^2 {\cal K}\,.
\end{equation}
It is the anti-linear operator of charge conjugation. ${\cal K}$ is the complex conjugation operator. We  define the {\it self/anti-self} charge-conjugate 4-spinors 
in the momentum space~\cite{DVA1995}:
\begin{eqnarray}
{\cal C}\lambda^{S,A} ({\bf p}) &=& \pm \lambda^{S,A} ({\bf p})\,,\\
{\cal C}\rho^{S,A} ({\bf p}) &=& \pm \rho^{S,A} ({\bf p})\,.
\end{eqnarray}
Thus,
\begin{equation}
\lambda^{S,A} (p^\mu)=\pmatrix{\pm i\Theta \phi^\ast_L ({\bf p})\cr
\phi_L ({\bf p})}\,,
\end{equation}
and
\begin{equation}
\rho^{S,A} ({\bf p})=\pmatrix{\phi_R ({\bf p})\cr \mp i\Theta \phi^\ast_R ({\bf p})}\,.
\end{equation}
$\phi_L$, $\phi_R$ can be boosted with the Lorentz transformation $\Lambda_{L,R}$ 
matrices.\footnote{Such definitions of 4-spinors differ, of course, from the original Majorana definition in x-representation:
\begin{equation}
\nu (x) = \frac{1}{\sqrt{2}} (\Psi_D (x) + \Psi_D^c (x))\,,
\end{equation}
$C \nu (x) = \nu (x)$ that represents the positive real $C-$ parity field operator. However, the momentum-space Majorana-like spinors 
open various possibilities for description of neutral  particles 
(with experimental consequences, see~\cite{Kirchbach}). For instance, "for imaginary $C$ parities, the neutrino mass 
can drop out from the single $\beta $ decay trace and 
reappear in $0\nu \beta\beta $, a curious and in principle  
experimentally testable signature for a  non-trivial impact of 
Majorana framework in experiments with polarized sources."}

The rest $\lambda$ and $\rho$ spinors are:\footnote{The choice of the helicity parametrization (\ref{ds}) for
${\bf p}\rightarrow {\bf 0}$ is doubtful in Ref.~\cite{AG}, and it leads to unremovable contradictions, in my opinion.}
\begin{eqnarray}
\lambda^S_\uparrow ({\bf 0}) &=& \sqrt{\frac{m}{2}}
\pmatrix{0\cr i \cr 1\cr 0}\,,\,
\lambda^S_\downarrow ({\bf 0})= \sqrt{\frac{m}{2}}
\pmatrix{-i \cr 0\cr 0\cr 1}\,,\,\\
\lambda^A_\uparrow ({\bf 0}) &=& \sqrt{\frac{m}{2}}
\pmatrix{0\cr -i\cr 1\cr 0}\,,\,
\lambda^A_\downarrow ({\bf 0}) = \sqrt{\frac{m}{2}}
\pmatrix{i\cr 0\cr 0\cr 1}\,,\,\\
\rho^S_\uparrow ({\bf 0}) &=& \sqrt{\frac{m}{2}}
\pmatrix{1\cr 0\cr 0\cr -i}\,,\,
\rho^S_\downarrow ({\bf 0}) = \sqrt{\frac{m}{2}}
\pmatrix{0\cr 1\cr i\cr 0}\,,\,\\
\rho^A_\uparrow ({\bf 0}) &=& \sqrt{\frac{m}{2}}
\pmatrix{1\cr 0\cr 0\cr i}\,,\,
\rho^A_\downarrow ({\bf 0}) = \sqrt{\frac{m}{2}}
\pmatrix{0\cr 1\cr -i\cr 0}\,.
\end{eqnarray}
Thus, in this basis the explicite forms of the 4-spinors of the second kind  $\lambda^{S,A}_{\uparrow\downarrow}
({\bf p})$ and $\rho^{S,A}_{\uparrow\downarrow} ({\bf p})$
are:
\begin{eqnarray}
\lambda^S_\uparrow ({\bf p}) &=& \frac{1}{2\sqrt{E_p+m}}
\pmatrix{ip_l\cr i (p^- +m)\cr p^- +m\cr -p_r},
\lambda^S_\downarrow ({\bf p})= \frac{1}{2\sqrt{E_p+m}}
\pmatrix{-i (p^+ +m)\cr -ip_r\cr -p_l\cr (p^+ +m)}\nonumber\\
\\
\lambda^A_\uparrow ({\bf p}) &=& \frac{1}{2\sqrt{E_p+m}}
\pmatrix{-ip_l\cr -i(p^- +m)\cr (p^- +m)\cr -p_r},
\lambda^A_\downarrow ({\bf p}) = \frac{1}{2\sqrt{E_p+m}}
\pmatrix{i(p^+ +m)\cr ip_r\cr -p_l\cr (p^+ +m)}\nonumber\\
\\
\rho^S_\uparrow ({\bf p}) &=& \frac{1}{2\sqrt{E_p+m}}
\pmatrix{p^+ +m\cr p_r\cr ip_l\cr -i(p^+ +m)},
\rho^S_\downarrow ({\bf p}) = \frac{1}{2\sqrt{E_p+m}}
\pmatrix{p_l\cr (p^- +m)\cr i(p^- +m)\cr -ip_r}\nonumber\\
\\
\rho^A_\uparrow ({\bf p}) &=& \frac{1}{2\sqrt{E_p+m}}
\pmatrix{p^+ +m\cr p_r\cr -ip_l\cr i (p^+ +m)},
\rho^A_\downarrow ({\bf p}) = \frac{1}{2\sqrt{E_p+m}}
\pmatrix{p_l\cr (p^- +m)\cr -i(p^- +m)\cr ip_r}.\nonumber
\\
\end{eqnarray}
As we showed $\lambda -$ and $\rho -$ 4-spinors are not the eigenspinors of the helicity. Moreover, 
$\lambda$ and $\rho$ are not (if we use the parity matrix 
$P=\pmatrix{0&1\cr 1&0}R$)  the eigenspinors of the parity, as opposed to the Dirac case.
The indices $\uparrow\downarrow$ should be referred to the chiral helicity 
quantum number introduced 
in the 60s, $\eta=-\gamma^5 h$, for $\lambda$ spinors.
While 
\begin{equation}
Pu_\sigma ({\bf p}) = + u_\sigma ({\bf p})\,,
Pv_\sigma ({\bf p}) = - v_\sigma ({\bf p})\,,
\end{equation}
we have
\begin{equation}
P\lambda^{S,A} ({\bf p}) = \rho^{A,S} ({\bf p})\,,
P \rho^{S,A} ({\bf p}) = \lambda^{A,S} ({\bf p})
\end{equation}
for the Majorana-like momentum-space 4-spinors on the first quantization level.
In this basis one has
\begin{eqnarray}
\rho^S_\uparrow ({\bf p}) \,&=&\, - i \lambda^A_\downarrow ({\bf p})\,,\,
\rho^S_\downarrow ({\bf p}) \,=\, + i \lambda^A_\uparrow ({\bf p})\,,\,\\
\rho^A_\uparrow ({\bf p}) \,&=&\, + i \lambda^S_\downarrow ({\bf p})\,,\,
\rho^A_\downarrow ({\bf p}) \,=\, - i \lambda^S_\uparrow ({\bf p})\,.
\end{eqnarray}
The analogs of the spinor normalizations (for $\lambda^{S,A}_{\uparrow\downarrow}
({\bf p})$ and $\rho^{S,A}_{\uparrow\downarrow} ({\bf p})$) are the following ones:
\begin{eqnarray}
\overline \lambda^S_\uparrow ({\bf p}) \lambda^S_\downarrow ({\bf p}) \,&=&\,
- i m \,,\quad
\overline \lambda^S_\downarrow ({\bf p}) \lambda^S_\uparrow ({\bf p}) \,= \,
+ i m \,,\quad\\
\overline \lambda^A_\uparrow ({\bf p}) \lambda^A_\downarrow ({\bf p}) \,&=&\,
+ i m \,,\quad
\overline \lambda^A_\downarrow ({\bf p}) \lambda^A_\uparrow ({\bf p}) \,=\,
- i m \,,\quad\\
\overline \rho^S_\uparrow ({\bf p}) \rho^S_\downarrow ({\bf p}) \, &=&  \,
+ i m\,,\quad
\overline \rho^S_\downarrow ({\bf p}) \rho^S_\uparrow ({\bf p})  \, =  \,
- i m\,,\quad\\
\overline \rho^A_\uparrow ({\bf p}) \rho^A_\downarrow ({\bf p})  \,&=&\,
- i m\,,\quad
\overline \rho^A_\downarrow ({\bf p}) \rho^A_\uparrow ({\bf p}) \,=\,
+ i m\,.
\end{eqnarray}
All other conditions are equal to zero.

The $\lambda -$ and $\rho -$ spinors are connected with the $u-$ and $v-$ spinors by the  following formula:
\begin{eqnarray}
\pmatrix{\lambda^S_\uparrow ({\bf p}) \cr \lambda^S_\downarrow ({\bf p}) \cr
\lambda^A_\uparrow ({\bf p}) \cr \lambda^A_\downarrow ({\bf p})\cr} = {1\over
2} \pmatrix{1 & i & -1 & i\cr -i & 1 & -i & -1\cr 1 & -i & -1 & -i\cr i&
1& i& -1\cr} \pmatrix{u_{+1/2} ({\bf p}) \cr u_{-1/2} ({\bf p}) \cr
v_{+1/2} ({\bf p}) \cr v_{-1/2} ({\bf p})\cr},\label{connect}
\end{eqnarray}
provided that the 4-spinors have the same physical dimension.\footnote{The change of 
the mass dimension of the field operator~\cite{AG} has no sufficient foundations because the Lagrangian can be constructed on
using the coupled Dirac equations, see Ref.~\cite{DVNCA}. After that one can play with $\sqrt{m}$
to reproduce all possible mathematical results, which may (or may not) answer to the physical reality.}

We construct the field operators on using the procedure above with $\lambda^S_\eta (p)$.
Thus, the difference with the last equality of the equation  (\ref{fo}) is that 1) instead of
$u_h (\pm p)$ we have $\lambda^S_\eta (\pm p)$; 2) possible change of the annihilation operators,
$a_h \rightarrow c_\eta$. Apart, one can make corresponding changes due to 
normalization factors. Thus, we should have 
\begin{equation}
\sum_{\eta=\pm 1/2}^{} \lambda^A_\eta (p) d_\eta^\dagger (p) = \sum_{\eta=\pm 1/2}^{} \lambda^S_\eta (-p) c_\eta (-p)\,.\label{dcop1}
\end{equation}
On using the same procedure as in the subsection (2.1), we find surprisingly:
\begin{equation}
d^\dagger_\eta (p) = -\frac{ip_y}{p}\sigma^y_{\eta\tau}  c_\tau (-p)\,,\quad
c_\eta (-p) = -\frac{ip_y}{p}\sigma^y_{\eta\tau}  d^\dagger_\tau (p)\,.
\end{equation}
The bi-orthogonal anticommutation relations are given in Ref.~\cite{DVA1995}. See other details in Ref.~\cite{DVNCA,DVMPLA}.
Concerning with the $P$,$C$ and $T$ properties of the corresponding states see Ref.~\cite{DVMPLA} in this model.

The above-mentioned contradiction may be related to the possibility of the conjugation which is different from that of Dirac.
Both in the Dirac-like case and the Majorana-like case ($c_\eta (p)= e^{-i\varphi} d_\eta (p)$) we have difficulties in the construction of field operators.

\section{The Spin-1.}

\subsection{The Standard Basis.}

We use the results of Refs.~\cite{Novozh,Wein,Dv-book} in this Section.
The polarization vectors of the standard basis are defined~\cite{Var}:
\begin{eqnarray}
&&\epsilon^\mu
({\bf 0}, +1)= -{1\over \sqrt{2}}\pmatrix{0\cr 1\cr i\cr 0 \cr}\,,\quad  
\epsilon^\mu ({\bf 0}, -1)= +{1\over
\sqrt{2}}\pmatrix{0\cr 1\cr -i\cr 0\cr}\,,\\
&&\epsilon^\mu ({\bf
0}, 0) = \pmatrix{0\cr 0\cr 0\cr 1\cr}\,,\quad
\epsilon^\mu ({\bf
0}, 0_t) = \pmatrix{1\cr 0\cr 0\cr 0\cr}\,.
\end{eqnarray}
The Lorentz transformations are ($\widehat p_i = p_i/\vert {\bf p}\vert$):
\begin{eqnarray} 
&& \epsilon^\mu ({\bf p}, \sigma) =
L^{\mu}_{\,\,\nu} ({\bf p}) \epsilon^\nu ({\bf 0},\sigma)\,,\\ 
&& L^{0}_{\,\,0} ({\bf p}) = \gamma\, ,\, L^{i}_{\,\,0} ({\bf p}) = 
L^{0}_{\,\,i} ({\bf p}) = \widehat p_i \sqrt{\gamma^2 -1}\, ,\,
L^{i}_{\,\,k} ({\bf p}) = \delta_{ik} + (\gamma -1) \widehat p_i \widehat
p_k \,.\nonumber\\ 
\end{eqnarray}
Hence, for the particles of the mass $m$ we have:
\begin{eqnarray} 
u^\mu
({\bf p}, +1)&=& -{N\over \sqrt{2}m}\pmatrix{-p^r\cr m+ {p^1 p^r \over
E_p+m}\cr im +{p^2 p^r \over E_p+m}\cr {p^3 p^r \over
E_p+m}\cr}\,,\quad 
u^\mu ({\bf p}, -1)= {N\over
\sqrt{2}m}\pmatrix{-p^l\cr m+ {p^1 p^l \over E_p+m}\cr -im +{p^2 p^l \over
E_p+m}\cr {p^3 p^l \over E_p+m}\cr}\,,\nonumber\\
&&\label{vp12}\\ u^\mu ({\bf
p}, 0) &=& {N\over m}\pmatrix{-p^3\cr {p^1 p^3 \over E_p+m}\cr {p^2 p^3
\over E_p+m}\cr m + {(p^3)^2 \over E_p+m}\cr}\,,\quad
u^\mu ({\bf p}, 0_t) = {N \over m} \pmatrix{E_p\cr -p^1
\cr -p^2\cr -p^3\cr}\,.
\end{eqnarray}
$N$ is the normalization constant for $u^\mu ({\bf p},\sigma)$. 
They are the eigenvectors of the parity operator ($\gamma_{00}= \mbox{diag} (1\,\, -1\,\, -1\,\, -1)$):
\begin{equation}
\hat P u_\mu (-{\bf p}, \sigma) = - u_\mu ({\bf p}, \sigma)\,,\quad
\hat P u_\mu (-{\bf p}, 0_t) = +u_\mu ({\bf p}, 0_t)\,.
\end{equation}
It is assumed that they form the complete orthonormalized system of the $(1/2,1/2)$
represntation, $\epsilon_\mu^\ast ({\bf p}, 0_t) \epsilon^\mu ({\bf p}, 0_t)=1$, $\epsilon_\mu^\ast ({\bf p}, \sigma^\prime ) \epsilon^\mu ({\bf p}, \sigma)=-\delta_{\sigma^\prime \sigma}$.

\subsection{The Helicity Basis.}
The helicity operator is:
\begin{equation}
{({\bf S}\cdot {\bf p})\over p} = {1\over p} \pmatrix{0&0&0&0\cr
0&0&-ip^3&ip^2\cr
0&ip^3&0&-ip^1\cr
0&-ip^2&ip^1&0\cr}\,,\,\,
{({\bf S}\cdot {\bf p})\over p} \epsilon^\mu_{\pm 1} = \pm \epsilon^\mu_{\pm 1}\,,\,\,{({\bf S}\cdot {\bf p})\over p} \epsilon^\mu_{0,0_t} = 0\,.
\end{equation}
The eigenvectors are:
\begin{eqnarray}
&&\epsilon^\mu_{+1}= {1\over \sqrt{2}} {e^{i\alpha}\over p} \pmatrix{0\cr
{-p^1 p^3 +ip^2 p\over \sqrt{(p^1)^2 +(p^2)^2}}\cr {-p^2 p^3 -ip^1
p\over \sqrt{(p^1)^2 +(p^2)^2}}\cr \sqrt{(p^1)^2 +(p^2)^2}\cr}\,,\,
\epsilon^{\mu }_{-1}={1\over \sqrt{2}}{e^{i\beta}\over p}
\pmatrix{0\cr {p^1 p^3 +ip^2 p\over \sqrt{(p^1)^2 +(p^2)^2}}\cr {p^2 p^3
-ip^1 p\over \sqrt{(p^1)^2 +(p^2)^2}}\cr -\sqrt{(p^1)^2 +(p^2)^2}\cr} \nonumber\\
&&\\
&&\epsilon^{\mu }_0={\frac{1}{m}}\pmatrix{p\cr {E \over p}
p^1 \cr {E \over p} p^2 \cr{E \over p} p^3\cr}\,,\quad
\epsilon^{\mu }_{0_{t}}={\frac{1}{m}}\pmatrix{E_p\cr p^1\cr
p^2\cr p^3\cr}\,.
\end{eqnarray}
The normalization is the same as in the standard basis.
The eigenvectors $\epsilon^\mu_{\pm  1}$ are not the eigenvectors
of the parity operator ($\gamma_{00} R$) of this representation. However, the $\epsilon^\mu_{1,0}$,
$\epsilon^\mu_{0,0_t}$
are. Surprisingly, the latter have no well-defined massless limit. In order to get the well-known massless limit one should use the basis of the light-front form reprersentation, cf.~\cite{Ahl-lf}.

\subsection{The Field Operators.} 

Various-type field operators are possible in this representation. Let us remind the procedure to get them.
Again, during the calculations below we have to present $1=\theta (p_0) +\theta (-p_0)$
in order to get positive- and negative-frequency parts.  Meanwhile, the Heaviside $\theta-$ function is  not defined in $p_0=0$. In general, due to integral theorems this presentation is possible even for distributions because we use the $\theta$- function in the integrand.\footnote{However, remember, that
we have the $p_0=0$ solution of the Maxwell equations. It has the experimental confirmation (for instance, the stationary magnetic field $curl {\bf B}=0$).}
\begin{eqnarray}
&&A_\mu (x) = {1\over (2\pi)^3} \int d^4 p \,\delta (p^2 -m^2) e^{-ip\cdot x}
A_\mu (p) =\nonumber\\
&=& {1\over (2\pi)^3} \sum_{\lambda}^{}\int d^4 p \delta (p_0^2 -E_p^2) e^{-ip\cdot x}
\epsilon_\mu (p,\lambda) a_\lambda (p) =\nonumber\\
&=&{1\over (2\pi)^3} \int {d^4 p \over 2E_p} [\delta (p_0 -E_p) +\delta (p_0 +E_p) ] 
[\theta (p_0) +\theta (-p_0) ]e^{-ip\cdot x}
A_\mu (p) =\nonumber\\
&=&{1\over (2\pi)^3} \int {d^4 p \over 2E_p} [\delta (p_0 -E_p) +\delta (p_0 +E_p) ] \left
[\theta (p_0) A_\mu (p) e^{-ip\cdot x}  + \right.\nonumber\\
&+&\left.\theta (p_0) A_\mu (-p) e^{+ip\cdot x} \right ]
={1\over (2\pi)^3} \int {d^3 {\bf p} \over 2E_p} \theta(p_0)  
[A_\mu (p) e^{-ip\cdot x}  + A_\mu (-p) e^{+ip\cdot x} ]
=\nonumber\\
&=&{1\over (2\pi)^3} \sum_{\lambda}^{}\int {d^3 {\bf p} \over 2E_p}   
[\epsilon_\mu (p,\lambda) a_\lambda (p) e^{-ip\cdot x}  + \epsilon_\mu (-p,\lambda) 
a_\lambda (-p) e^{+ip\cdot x} ]\,.
\end{eqnarray}
We should transform the second part to $\epsilon_\mu^\ast (p,\lambda) b_\lambda^\dagger (p)$ as usual. In such a way we obtain the states which are considered to be the charge-conjugate 
states. In this Lorentz group representation the charge conjugation operator is just the complex conjugation operator for 4-vectors. Of course, one can try to get $P$-conjugates or $CP$-conjugate states too. 
We postulate
\begin{equation}
\sum_{\lambda}^{} \epsilon_\mu (-p,\lambda) a_\lambda (-p) = 
\sum_{\lambda}^{} \epsilon_\mu^\ast (p,\lambda) b_\lambda^\dagger (p)\,.
\label{expan}
\end{equation}
Then we multiply both parts by $\epsilon^\mu (p,\sigma)$, and use the normalization conditions for polarization vectors.

In the $({1\over 2}, {1\over 2})$ representation we can also expand
(apart of the equation (\ref{expan})) in a different way. For example,
\begin{equation}
\sum_{\lambda}^{} \epsilon_\mu (-p, \lambda) c_\lambda (-p) =
\sum_{\lambda}^{} \epsilon_\mu (p, \lambda) d^\dagger_\lambda (p)\,.
\end{equation}
From the first definition we obtain:
\begin{eqnarray}
&&\pmatrix{b^\dagger_{0_t} (p)\cr
-b^\dagger_{+1} (p)\cr
-b^\dagger_{0} (p)\cr
-b^\dagger_{-1} (p)\cr} =
 \sum_{\mu\lambda}^{} \epsilon^\mu (p,\sigma) 
\epsilon_\mu (-p,\lambda) a_\lambda (-p)= \sum_{\lambda} \Lambda_{\sigma\lambda}^{(1a)} a_\lambda (-p)=\nonumber\\ 
&=&\pmatrix{-1&0&0&0\cr
0&\frac{p_r^2}{p^2}& - \frac{\sqrt{2} p_z p_r}{p^2}& \frac{p_z^2}{p^2}\cr
0&- \frac{\sqrt{2} p_z p_r}{p^2}&-1 +\frac{2p_z^2}{p^2}& +\frac{\sqrt{2}p_z p_l}{p^2}\cr
0&\frac{p_z^2}{p^2}&+ \frac{\sqrt{2} p_z p_l}{p^2}& \frac{p_l^2}{p^2}\cr}\pmatrix{a_{00} (-p)\cr a_{11} (-p)\cr
a_{10} (-p)\cr a_{1-1} (-p)\cr}\,.
\end{eqnarray}
From the second definition $\Lambda^{1b}_{\sigma\lambda} =  \sum_{\mu} \epsilon^{\mu^\ast} (p, \sigma) \epsilon_\mu (-p, \lambda)$ 
we have
\begin{eqnarray}
&&\pmatrix{d^\dagger_{0_t} (p)\cr
-d^\dagger_{+1} (p)\cr
-d^\dagger_{0} (p)\cr
-d^\dagger_{-1} (p)\cr} =
 \sum_{\mu\lambda}^{} \epsilon^{\mu^\ast} (p,\sigma) 
\epsilon_\mu (-p,\lambda) c_\lambda (-p)= \sum_{\lambda} \Lambda_{\sigma\lambda}^{(1b)} c_\lambda (-p)=\nonumber\\ 
&=&\pmatrix{-1&0&0&0\cr
0&\frac{-p_z^2}{p^2}& - \frac{\sqrt{2} p_z p_l}{p^2}& \frac{-p_l^2}{p^2}\cr
0&-\frac{\sqrt{2} p_z p_r}{p^2}&-1 +\frac{2p_z^2}{p^2}& +\frac{\sqrt{2}p_z p_l}{p^2}\cr
0&-\frac{p_r^2}{p^2}&+ \frac{\sqrt{2} p_z p_r}{p^2}& -\frac{p_z^2}{p^2}\cr}\pmatrix{c_{00} (-p)\cr c_{11} (-p)\cr
c_{10} (-p)\cr c_{1-1} (-p)\cr}=-g^{\sigma\lambda} +({\bf S}\cdot {\bf n})^2_{\sigma\lambda}\,.\nonumber\\
&&
\end{eqnarray}
Possibly, we should think about modifications of the Fock space in this case. Alternatively, one can think to introduce several field operators for the $({1\over 2}, {1\over 2})$ representation. The Majorana-like anzatz is compatible for the $0_t$ polarization state\footnote{$00$ in other notation.} only in this basis of this representation.

However, the corresponding matrices $\Lambda^2$ in the helicity basis are different. Here they are:
\begin{eqnarray}
&&\pmatrix{b^\dagger_{0_t} (p)\cr
-b^\dagger_{+1} (p)\cr
-b^\dagger_{0} (p)\cr
-b^\dagger_{-1} (p)\cr} =
 \sum_{\mu\lambda}^{} \epsilon^\mu (p,\sigma) 
\epsilon_\mu (-p,\lambda) a_\lambda (-p)= \sum_{\lambda} \Lambda_{\sigma\lambda}^{(2a)} a_\lambda (-p)=\nonumber\\ 
&=&-\pmatrix{1&0&0&0\cr
0&e^{2i\alpha}&0&0\cr
0&0&1&0\cr
0&0&0&e^{2i\beta}}\pmatrix{a_{00} (-p)\cr a_{11} (-p)\cr
a_{10} (-p)\cr a_{1-1} (-p)\cr}\,,
\end{eqnarray}
and
\begin{eqnarray}
&&\pmatrix{d^\dagger_{0_t} (p)\cr
-d^\dagger_{+1} (p)\cr
-d^\dagger_{0} (p)\cr
-d^\dagger_{-1} (p)\cr} =
 \sum_{\mu\lambda}^{} \epsilon^{\mu ^\ast} (p,\sigma) 
\epsilon_\mu (-p,\lambda) c_\lambda (-p)= \sum_{\lambda} \Lambda_{\sigma\lambda}^{(2b)} c_\lambda (-p)=\nonumber\\ 
&=&-\pmatrix{1&0&0&0\cr
0&0&0&-e^{-i(\alpha-\beta)}\cr
0&0&1&0\cr
0&-e^{+i(\alpha-\beta)}&0&0\cr}\pmatrix{c_{00} (-p)\cr c_{11} (-p)\cr
c_{10} (-p)\cr c_{1-1} (-p)\cr}\,.
\end{eqnarray}
This is compatible with the Majorana-like anzatzen.

Of course, the same procedure can be applied in the construction of the quantum field operator for $F_{\mu\nu}$.

\subsection{The $(1,0)\oplus (0,1)$ Representation.}

The solutions of the Weinberg-like equation
\begin{equation}
[\gamma^{\mu\nu} \partial_\mu \partial_\nu - {(i\partial/\partial t)\over E} 
m^2 ] \Psi (x) =0\,.
\end{equation}
are found in Refs.~\cite{Sankar,Novozh,DVA,DVO94}. Here they are:
\begin{eqnarray}
&&u_\sigma ({\bf p})= \pmatrix{D^S (\Lambda_R) \xi_\sigma ({\bf 0})\cr 
D^S (\Lambda_L)\xi_\sigma ({\bf 0})\cr}\,,\,\,
v_\sigma ({\bf p})=\pmatrix{D^S (\Lambda_R \Theta_{[1/2]}) \xi_\sigma^\ast ({\bf 0})\cr 
- D^S (\Lambda_L \Theta_{[1/2]})\xi_\sigma^\ast ({\bf 0})\cr}=\Gamma^5 u_\sigma ({\bf p}),\nonumber\\ 
&&\\
&&\Gamma^5 = \pmatrix{1_{3\times 3}&0_{3\times 3}\cr
0_{3\times 3}&-1_{3\times 3}\cr}\,,
\end{eqnarray}
where $D^S$ is the matrix of the $(S,0)$ representation of the spinor group $SL (2,c)$. 

In the $(1,0)\oplus (0,1)$ representation 
the procedure of derivation of the creation operators leads to somewhat different situation:
\begin{equation}
\sum_{\sigma=0,\pm 1}^{} v_\sigma (p) b_\sigma^\dagger (p) = \sum_{\sigma=0,\pm 1}^{} u_\sigma (-p) a_\sigma (-p)\,,\quad 
\mbox{hence}\,\,b_\sigma^\dagger (p) = 0\,.
\end{equation}
However, if we return to the original Weinberg equations $[\gamma^{\mu\nu} \partial_\mu \partial_\nu \pm 
m^2 ] \Psi_{1,2} (x) =0$ with the field operators:
\begin{eqnarray}
\Psi_1 (x) &=& \frac{1}{(2\pi)^3}\sum_\mu \int \frac{d^3 {\bf p}}{2E_p} [ u_\mu ({\bf p}) a_\mu ({\bf p}) e^{-ip_\mu\cdot x^\mu}
+ u_\mu ({\bf p}) b_\mu^\dagger ({\bf p}) e^{+ip_\mu\cdot x^\mu}],\\
\Psi_2 (x) &=& \frac{1}{(2\pi)^3}\sum_\mu \int \frac{d^3 {\bf p}}{2E_p} [ v_\mu ({\bf p}) c_\mu ({\bf p}) e^{-ip_\mu\cdot x^\mu}
+ v_\mu ({\bf p}) d^\dagger_\mu ({\bf p}) e^{+ip_\mu\cdot x^\mu}],
\end{eqnarray}
we obtain
\begin{eqnarray}
b_\mu^\dagger (p) &=& [1-2({\bf S}\cdot {\bf n})^2]_{\mu\lambda} a_\lambda (-p)\,,\label{ba}\\
d_\mu^\dagger (p) &=& [1-2({\bf S}\cdot {\bf n})^2]_{\mu\lambda} c_\lambda (-p)\,\label{ca}.
\end{eqnarray}
The applications of $\overline u_\mu (-p) u_\lambda (-p) = \delta_{\mu\lambda}$ and 
$\overline u_\mu (-p) u_\lambda (p) = [1- 2({\bf S}\cdot {\bf n})^2]_{\mu\lambda}$ prove that the equations
are self-consistent (similarly to the subsection (2.1)).
This situation signifies that in order to construct the Sankaranarayanan-Good field operator (which was used by Ahluwalia, Johnson and 
Goldman~\cite{DVA}) we need additional postulates. One can try to construct 
the left- and the right-hand side of the field operator separately each other.
In this case the commutation relations may also be more complicated.

Is it possible to apply the Majorana-like anzatz to the $(1,0)+(0,1)$ fields? Repeating the procedure of the Section 2.3, 
on using (\ref{ba}) and the Majorana posulate we come to:
\begin{equation}
a_\mu^\dagger (p) = + e^{+i\varphi} [1-2({\bf S}\cdot {\bf n})^2]_{\mu\lambda} a_\lambda (-p)\,.\label{ma1S1}
\end{equation}
On the other hand, on using the inverse relation, namely, that for $a_\mu (-p)$, we make the substitutions $E_p \rightarrow -E_p$, ${\bf p} 
\rightarrow -{\bf p}$ to obtain 
\begin{equation}
a_\mu (p) = + [1-2({\bf S}\cdot {\bf n})^2]_{\mu\lambda} b_\lambda^\dagger (-p)\,.\label{ma2S1}
\end{equation}
The totally reflected Majorana anzatz is $b_\mu (-E_p, -{\bf p}) = e^{i\varphi} a_\mu (-E_p, -{\bf p})$. Thus,
\begin{equation}
b_\mu^\dagger (- p) = e^{-i\varphi} a_\mu^\dagger (- p)\,.
\end{equation}
Combining with (\ref{ma2S1}),  we come to
\begin{equation}
a_\mu (p) = + e^{-i\varphi} [1-2({\bf S}\cdot {\bf n})^2]_{\mu\lambda} a_\lambda^\dagger (-p)\,,
\end{equation}
and 
\begin{equation}
a_\mu^\dagger (p) = +e^{+i\varphi} [1-2({\bf S}^\ast \cdot {\bf n})^2]_{\mu\lambda} a_\lambda (-p)\,.
\end{equation}
In the basis where $S_z$ is diagonal the matrix $S_y$ is imaginary~\cite{Var}. So, $({\bf S}^\ast \cdot {\bf n})= S_x n_x - S_y n_y + S_z n_z$, and $({\bf S}^\ast\cdot {\bf n})^2 \neq ({\bf S}\cdot {\bf n})^2$ in the case of $S=1$. So, we conclude that there is the same problem in this 
point, in the aplication of the Majorana-like anzatz, as in the case of spin-1/2. Similarly, one can proceed with (\ref{ca}).

Meanwhile, the attempts of constructing the self/anti-self charge conjugate states failed in Ref.~\cite{DVA1995}. Instead, the $\Gamma^5 S^c_{[1]}-$ self/anti-self conjugate states
have been constructed therein.


\section{Conclusions.}

We conclude that something is missed in the foundations of both the original Majorana theory and its generalizations.
Similar problems exist in the theories of higher spins.

\medskip

{\bf Acknowledgements.} I acknowledge discussions with colleagues at recent conferences.
I am grateful to the Zacatecas University for professorship. 


\smallskip

\begin{thebibliography}{99}

\footnotesize{

\bibitem{DV1} V. V. Dvoeglazov, Hadronic J. Suppl. {\bf 18}, 239 (2003).\\[-6mm]

\bibitem{DV2} V. V. Dvoeglazov, Int. J. Mod. Phys. B{\bf 20}, 1317 (2006).\\[-6mm]

\bibitem{DV3}  V. V. Dvoeglazov, in {\it Einstein and Others: Unification.} Ed. V. V. Dvoeglazov. (Nova Sci. Pubs., Hauppauge, NY, USA, 2015), p. 211.\\[-6mm]

\bibitem{DV4} V. V. Dvoeglazov, Z. Naturforsch. A {\bf 71}, 345 (2016). \\[-6mm]

\bibitem{DVCONF1} V. V. Dvoeglazov, SFIN {\bf 22}-A1, 157 (2009); J. Phys. Conf. Ser. {\bf 284}, 012024 (2011); Bled Workshops {\bf 10}-2, 52 (2009); ibid. {\bf 11}-2, 9 (2010); in {\it Beyond the Standard Model. Proceedings of the Vigier Symposium.} (Noetic Press, Orinda, USA, 2005), p. 388.\\[-6mm]

\bibitem{DVCONF2} V. V. Dvoeglazov, Bled Workshops {\bf 14}-2, 199 (2013).\\[-6mm]

\bibitem{Dirac} P. A. M. Dirac, Proc. Roy. Soc. Lond. A{\bf 117}, 610 (1928).\\[-6mm] 

\bibitem{Sakurai}  J. J. Sakurai, {\it Advanced Quantum Mechanics.} (Addison-Wesley, 1967).\\[-6mm]

\bibitem{Ryder} L. H. Ryder, {\it Quantum Field Theory.} (Cambridge University Press, Cambridge, 1985).\\[-6mm]

\bibitem{Itzyk} C. Itzykson and J.-B. Zuber, {\it Quantum Field Theory}
(McGraw-Hill Book Co., 1980).\\[-6mm]

\bibitem{Bogoliubov} N. N. Bogoliubov and D. V. Shirkov, {\it Introduction to 
the Theory of Quantized Fields.} 2nd Edition. (Nauka, Moscow, 1973).\\[-6mm]

\bibitem{Var} D. A. Varshalovich, A. N. Moskalev and V. K. Khersonski\u{\i},
{\it Quantum Theory of Angular Momentum} (World Scientific, Singapore, 1988), \S 6.2.5.\\[-6mm]

\bibitem{Dv1} V. V. Dvoeglazov, Fizika B{\bf 6}, 111 (1997).\\[-6mm]

\bibitem{DVIJTP} V. V. Dvoeglazov, Int. J. Theor. Phys. {\bf 43}, 1287 (2004).\\[-6mm]

\bibitem{BLP} V. B. Berestetski\u{\i}, E. M. Lifshitz and L. P.  Pitaevski\u{\i}, {\it Quantum
Electrodynamics.} (Pergamon Press, 1982, translated from the Russian), \S
16.\\[-6mm]

\bibitem{extra} V. V. Dvoeglazov, Nuovo Cim. B{\bf 111}, 483 (1996).\\[-6mm]

\bibitem{DVA1995} D. V. Ahluwalia, Int. J. Mod. Phys. A{\bf 11}, 1855 (1996).\\[-6mm]

\bibitem{Wein} S. Weinberg, {\it The Quantum Theory of Fields. Vol. I.
Foundations.} (Cambridge University Press, Cambridge, 1995).\\[-6mm]

\bibitem{Greinb} W. Greiner, {\it Field Quantization.} (Springer, 1996).\\[-6mm] 

\bibitem{Tokuoka} Z. Tokuoka, Prog. Theor. Phys. {\bf 37},  603 (1967).\\[-6mm]

\bibitem{GR} H. M. R\"uck and W. Greiner, J. Phys. G: Nucl. Phys. {\bf 3},
657 (1977).\\[-6mm]

\bibitem{Majorana} E. Majorana, Nuovo Cim. {\bf 14}, 171 (1937).\\[-6mm]


\bibitem{Kirchbach} M. Kirchbach, C. Compean and L. Noriega, Eur. Phys. J. A{\bf 22}, 149 (2004).\\[-6mm]

\bibitem{AG} D. V. Ahluwalia and D. Grumiller, JCAP {\bf 0507}, 012 (2005).\\[-6mm]

\bibitem{DVNCA} V. V. Dvoeglazov, Nuovo Cim. A{\bf 108}, 1467 (1995).\\[-6mm]


\bibitem{DVMPLA} V. V. Dvoeglazov, Mod. Phys. Lett. A{\bf 12}, 2741 (1997); J. Phys. Conf. Ser. {\bf 1010}, 012011 (2018).\\[-6mm]

\bibitem{Novozh} Yu. V. Novozhilov, {\it Introduction to Elementary Particle
Physics} (Pergamon Press, 1975).\\[-6mm]

\bibitem{Dv-book} V. V. Dvoeglazov, in {\it Photon: Old Problems in Light of New Ideas.} Ed. V. V. Dvoeglazov  
(Nova Sci. Pubs., Huntington, NY, USA, 
2000).\\[-6mm]

\bibitem{Ahl-lf} D. V. Ahluwalia and M. Sawicki, Phys. Rev. D{\bf 47}, 5161 (1993); Phys. Lett. B{\bf 335}, 24 (1994).\\[-6mm]

\bibitem{Sankar} A. Sankaranarayanan and R. H. Good, Jr., Nuovo Cim. {\bf 36}, 1303 (1965).\\[-6mm]

\bibitem{DVA} D. V. Ahluwalia, M. B. Johnson and T. Goldman, Phys. Lett. B{\bf 316}, 102 (1993).\\[-6mm]

\bibitem{DVO94} V. V. Dvoeglazov, Int. J. Theor. Phys. {\bf 37}, 1915 (1998).\\[-6mm]


}


\end{thebibliography}
\end{document}